\documentclass[a4paper,11pt]{article}
\usepackage{amsmath}
\usepackage{amsfonts}
\usepackage{amssymb}
\usepackage{graphicx,epsfig}
\usepackage{hyperref}
\begin{document}
\title{On Spectrum Generating Algebra of the Heun Operator}
\author{Priyasri Kar$^a$, Ritesh K. Singh$^b$, Ananda Dasgupta$^c$, 
Prasanta K. Panigrahi$^d$\\[0.2cm]
{\it Department of Physical Sciences,}\\ {\it Indian Institute of Science Education and Research Kolkata, Mohanpur 741246, India}\\[0.2cm]
$^a$ {\tt pk12rs063@iiserkol.ac.in},
$^b$ {\tt ritesh.singh@iiserkol.ac.in},\\
$^c$ {\tt adg@iiserkol.ac.in},
$^d$ {\tt pprasanta@iiserkol.ac.in}
}
\maketitle
\begin{abstract}
The Heun operator has been cast, in terms of the elements of an underlying
$su(1,1)$ algebra, under specific parametric conditions, for the purpose of
spectrum generation.
These elements are differential operators of \emph{degrees} $\pm 1/2$ and $0$.
 It is found that the regular singularities at $0$ and
$\infty$ of the general Heun equation must be \emph{elementary} under the required
parametric conditions. The spectrum generation has been demonstrated through a set of examples.
\end{abstract}

Differential equations, in general, \emph{do not} admit exact 
solutions. Several techniques have been devised, to obtain partial, or in 
certain cases, the complete eigenspace of the differential operator under 
study~\cite{willard_miller}. In this regard, 
the well known factorization method, which at a fundamental level reduces the 
order of the differential operator involved, has found extensive 
applications~\cite{hull_infeld}. The harmonic oscillator spectrum 
generating Heisenberg algebra, the Darboux transformation, factorization of 
the hypergeometric equation by 
Schr\"{o}dinger~\cite{schroedinger} and subsequent study of a large class of 
equations by Infeld and Hull~\cite{hull_infeld} serve as classic examples. 
Supersymmetric quantum mechanics (SUSYQM) uses factorization, to reveal 
unexpected interconnections amongst apparently different quantal 
problems~\cite{khare_cooper_sukhatme_PR,khare_cooper_sukhatme_book,asim_book}.

Group theoretical structures also have played significant role in unravelling
the symmetry of differential operators~\cite{willard_miller}. A partial 
algebraization of the whole
solution space has been achieved~\cite{turbiner,shif_tur,ush_book} for a wide
class of quantal quasi-exactly solvable (QES) Hamiltonians of one or more
dimensions, which are essentially hermitian differential operators of the second
order. These Hamiltonians, or their appropriately similarity transformed
versions, often take the form of some well known differential operators, e.g.,
Heun, Lam\'{e} etc. Hence, the study of the underlying symmetries of these
operators is of immense importance and several attempts have been made to this
end. For example, the Lam\'{e} equation has 
been expressed as a bilinear in both $su(2)$~\cite{alhassid} 
and $su(1,1)$~\cite{kus&li} generators. 
The latter case relates the non-unitary representations of $su(1,1)$, to the
eigenfunction of the periodic Lam\'{e} potential. 
Closed form solutions for confluent hypergeometric
and hypergeometric equations have been found~\cite{pkp_gurappa_sigma}, 
using a connection between
the space of monomials and the solution space of the above 
equations~\cite{pkp_gurappa_2000, pkp_gurappa_2004}, which
reveals the underlying $su(1,1)$ and deformed $sl(2)$ structures, 
respectively~\cite{pkp_shreecharan}.

%
%
 
Recently, the Heun operator
has been cast~\cite{Arunesh} in terms of the elements of cubic deformations of
$sl(2)$ algebra.  This has led to two known solutions of a Heun equation,
encountered earlier by Christ and Lee \cite{LeeChrist}. However, non-linear
deformations of $sl(2)$, are associated with non-trivial representation theory
\cite{rocek}. 
Casting the differential operator in terms of linear $sl(2)$ algebra enables one to exploit available representation theory. Linear $su(1,1)$ symmetry of the Heun operator is present in the literature~\cite{turbiner}. In the present work, we use a \emph{different} representation of $su(1,1)$ generators. Under certain parametric conditions, this reveals a new symmtery of the equation and leads to a number of solutions, that are unavailable from the methods in \cite{turbiner}. It is found that this new symmetry exists if the singularities 
at $z=0$ and $z=\infty$ are 
\emph{elementary}\footnote{Any regular singularity can be characterized by two
exponents $\rho_1$ and $\rho_2$, which are the two roots of the indicial
equation. The cases with $|\rho_1-\rho_2|=1/2$ are known as elementary
singularity and are of special significance. All regular and irregular
singularities can be obtained by the coalescence of two and three or more
elementary singularities, respectively~\cite{ince}.}. Under these conditions,
an exact map is established between the Heun solution space and the 
unitary and non-unitary representations of $su(1,1)$ algebra.

The Heun equation is given by \cite{ronveauxbook}
\begin{equation}\label{heuneqn}
\mathcal{H}y(z)=\frac{d^2y(z)}{dz^2}+\left(\frac{\gamma}{z}+\frac{\delta}{z-1}+\frac{\varepsilon}{z-a}\right)\frac{dy(z)}{dz}+\frac{\alpha\beta
z-q}{z(z-1)(z-a)}y(z)=0,
\end{equation}
with regular singularities at $z$ = $0, \ 1, \ a(\ne 0,1) \mbox{ and} \ \infty$;
the exponents at these singularities being $(0, \ 1-\gamma)$, $(0,\ 1-\delta)$, 
$(0, \ 1-\varepsilon)$ and $(\alpha, \ \beta)$, respectively. Here, $q$ plays 
the role of the \emph{eigen parameter}. In this work, we study Heun equations with 
only real parameters. The equation, being a second order linear 
homogeneous equation with $4$ regular singularities, satisfies the Fuchsian condition: 
\begin{equation}\label{fuchsiancondn}
\gamma+\delta+\varepsilon=\alpha+\beta+1,
\end{equation}
allowing elimination of $\varepsilon$ in favor of the others. 
Eq.~(\ref{heuneqn}) can be written as:
\begin{equation}\label{gendiffeqn}
\mathcal{H}y(z)=\left[f_{1}(z)\frac{d^2}{dz^2}+f_2(z)\frac{d}{dz}+f_3(z)\right]y(z)
= 0,
\end{equation}where, $f_{1}(z)=a_{0}z^3+a_{1}z^2+a_{2}z$, $\space$ $f_{2}(z)=a_{3}z^2+a_{4}z+a_{5}$ $\space$ and $\space$ $f_{3}(z)=a_{6}z+a_{7}$.
The parameters $a_{i} \in \mathbb{R}, \ \mbox{for} \ i = 0,\ldots,7 $ are 
given by,
\begin{subequations}\label{a0toa7}
\begin{align}
a_0 =& 1 \qquad \qquad \qquad a_3 = 1+\alpha+\beta \qquad \qquad \qquad \qquad
\qquad a_6 = \alpha\beta \label{a0a3a6} \\
a_1 =& -(a+1) \hspace{.9cm} a_4 = -[a\gamma+a\delta-\delta+\alpha+\beta+1]
\hspace{.9cm} a_7 = -q \label{a1a4a7}\\
a_2 =& a \qquad \qquad \qquad a_5 =  a\gamma. \label{a2a5} 
\end{align}
\end{subequations}
Evidently, the Heun equation consists of differential operators of 
{\em degrees}\footnote{The degree $d$, of an operator $O_d$, is defined as the
change in the power of a monomial, when acted upon by it, i.e., $O_d \ z^p
\propto z^{p+d}$. } 
$+1,\ 0$ and $-1$, which are denoted by $P_+$, $F(P_0)$ and $P_-$, respectively. Eq.~(\ref{gendiffeqn}) can then be rewritten 
as~\cite{Arunesh},
\begin{equation}
\mathcal{H}y(z)=[P_+ + F(P_0) + P_-] y(z) = 0 
\label{heunclass}
\end{equation}
where,
\begin{eqnarray}
P_{+} = a_{0}z^3\frac{d^2}{dz^2}+a_{3}z^2\frac{d}{dz}+a_{6}z, \quad 
P_{0} = z\frac{d}{dz}-j,\quad 
P_{-}=a_{2}z\frac{d^2}{dz^2}+a_{5}\frac{d}{dz} \\ 
\mbox{and} \quad F(P_0)=a_{1}P_0^2+\left((2j-1)a_1+a_4\right)P_0
+\left(j(j-1)a_1+ja_4+a_7\right).
\end{eqnarray}
The above operators satisfy cubic deformation of $sl(2)$ algebra among 
themselves, which has been exploited \cite{Arunesh} to tract a part of the 
eigenspace of a Heun operator analytically, as mentioned earlier.

With the aim to cast the Heun operator in terms of linear $su(1,1)$ algebra, we
define the \emph{constituent} operators as,
\begin{eqnarray}\label{boperators}
E_+ = 2 z^{3/2}\frac{d}{dz}+2\mu\sqrt{z} \label{bplus}, \hspace{1cm}
H = 2 z\frac{d}{dz}+\mu+\nu \label{bzero} \quad \mbox{and} \quad
E_- = 2 z^{1/2}\frac{d}{dz}+\frac{2\nu}{\sqrt{z}}; \label{bminus}
\end{eqnarray}
where $\mu$, $\nu$ and $\lambda$ are parameters. The operators $E_+$,
$H$ and $E_-$, have degrees $+1/2, \ 0$ and $-1/2$, respectively. They are found to satisfy $su(1,1)$ algebra, with
commutators,
\begin{eqnarray}\label{bcommutators}
[H,E_{\pm}]=\pm E_{\pm}, \hspace{1cm} [E_+,E_-]=-2H
\end{eqnarray}
and the Casimir:
\begin{equation}\label{unshifted_casimir}
C(\mu,\nu)=\dfrac{1}{2}\left(E_+E_-+E_-E_+\right)-H^2=-(\mu-\nu)(\mu-\nu-1).
\end{equation}
The problem in question, determines the parameters of the
constituent operators, which determine the Casimir value.
The Casimir value, in turn, determines the representations available for the algebra
associated with the problem.

\begin{figure}
\includegraphics[width=16.5cm]{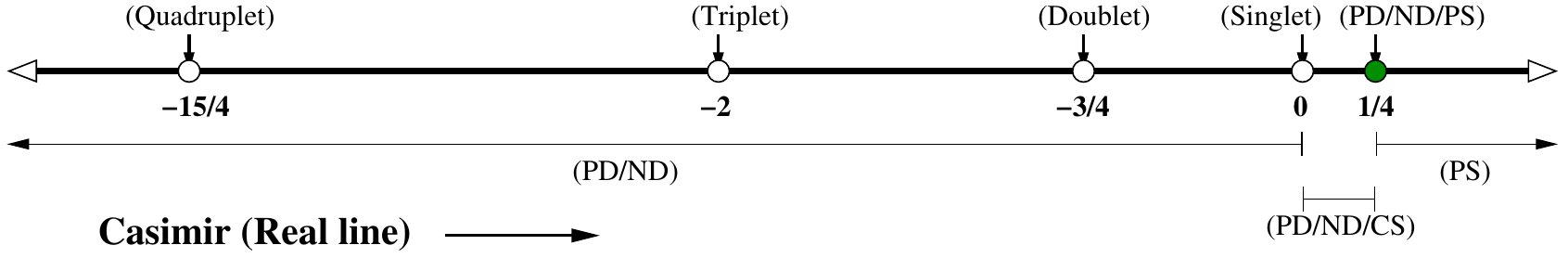}
\caption{\label{fig:rep}\emph{Casimir value for $su(1,1)$ and corresponding
representations are listed: positive discrete (PD), negative discrete (ND),
principal series (PS), complementary series (CS) and some finite dimensional
ones (open circles).}}
\end{figure}

We now proceed to construct the differential operators $P_+, P_-$ and $F(P_0)$,
from the constituent $su(1,1)$ generators and identify the conditions, under 
which this is possible. The constituent operators are of degrees $\pm1/2$ and 
$0$, whereas, the degrees of operators $P_{\pm}$ and $F(P_0)$ are $\pm1$ and 
$0$, respectively. Hence, it is clear that the complete Heun operator will 
comprise of linear and quadratic forms of the constituent 
operators\footnote{Cubic and
quartic combinations, such as $E_+E_+H$, $E_+E_+HH$, $E_+E_+E_+E_-$ (and similar
terms with $E_+$ and $E_-$ interchanged), are also of the desired degrees and
therefore could be used in the construction of the Heun operators. However, they contain
differential operators of higher orders, not present in the Heun equation. Thus,
the coefficients of those undesired terms must be put to zero, which revealed
no more parametric freedom than the chosen combinations.}.
Assuming the forms,
\begin{equation}
P_+ = c_+ E_+E_+ \qquad \mbox{and}  \qquad P_- = c_- E_-E_-
\end{equation}
where $c_{\pm}$ are constants, we obtain,
\begin{eqnarray}
&& a_0 = 4c_+, \ \ \ 
a_3 = 2c_+(3+4\mu)=\dfrac{a_0}{2}(3+4\mu),  \ \ 
a_6 = 2c_+\mu(1+2\mu)= \dfrac{a_0}{2}\mu(1+2\mu),\label{a0a4a7}\\
&& a_2 = 4c_-, 
\quad a_5 =  2c_-(1+4\nu)=\dfrac{a_2}{2}(1+4\nu), \quad
2c_-\nu(2\nu-1) = 0.\label{a6a2}
\end{eqnarray}
Using Eqs.~(\ref{a0a4a7}) and (\ref{a0a3a6}), one can solve for $\mu$, $\alpha$,
$\beta$:
\begin{equation}\label{factor_condn1}
|\alpha-\beta|=\dfrac{1}{2},
\end{equation}
implying the singularity at $z=\infty$ is elementary.
Similarly, using Eqs.~(\ref{a6a2}) and (\ref{a2a5}), one is left with 
just two values of $\nu$:
\begin{equation}\label{factor_condn2}
\nu=0 \implies a_5/a_2=\gamma=1/2 \qquad \mbox{or,} \qquad \nu=1/2 \implies 
a_5/a_2=\gamma=3/2.
\end{equation}
Above solutions of $\gamma$ imply that the singularity at $z=0$ is also 
elementary. Finally, $F(P_0)$ can be written in the form 
\begin{equation}
F(P_0) = c_2HH+c_1H+c_0,
\end{equation}
where,
\begin{equation}\label{a1a5a8}
 a_1 = 4c_2, \quad 
 a_4 = 2(c_1+2c_2(1+\mu+\nu)) \quad 
\mbox{and} \quad a_7 =-q=
c_0+(\mu+\nu)(c_1+c_2(\mu+\nu)).
\end{equation}
Thus, a Heun operator with elementary singularities at $z=0$ and $z=\infty$,
i.e., with parametric conditions Eqs.~(\ref{factor_condn1}) and
(\ref{factor_condn2}), becomes,
\begin{equation}\label{factorized_Heun}
\mathcal{H}=c_+E_+E_+ + c_-E_-E_- + c_2HH + c_1H + c_0.
\end{equation}
The choices of $E_+$ and $E_-$ fix the values of $\mu$ and $\nu$, respectively,
which in turn determines the Casimir, $C=C(\mu,\nu)$.
Finding the solution to Heun equation, then boils down to identifying a suitable
representation space $V_{C}$ of $su(1,1)$ and finding a 
$y\in V_{C}$, such that, $\mathcal{H}y=0$. 
For $su(1,1)$ algebra, there are five different classes of
representation spaces \cite{groenevelt} determined by the values of $C$ and $h$, the eigen value 
of operator $H$. These are shown in Fig.~\ref{fig:rep} and are listed below.
\begin{description}
\item{\bf Principal series (PS):} Infinite dimensional space,  
$C\ge 1/4$, $h\in(-\infty,\infty)$ and $(C,h)\neq(1/4,1/2)$,
\item{\bf Complementary series (CS):} Infinite dimensional space, $C\in
(0,1/4)$,  $h\in(-\infty,\infty), h\neq1/2$,
\item{\bf Positive discrete (PD):} Infinite dimensional space, $C\le 1/4$,
$h\ge0$, 
\item{\bf Negative discrete (ND):} Infinite dimensional space, $C\le 1/4$,
$h\le0$,
\item {\bf Finite dimensions:} Finite dimensional space,
$C=\frac{-(n^2-1)}{4}$, $h \in \left[\frac{-(n-1)}{2},\frac{(n-1)}{2}\right]$ 
with $n\in \mathbb{N}$.
\end{description}
For a given value of Casimir $C$, one of the available representation space 
$V_{C}$ can be chosen. The elements of this space are
$|C,h\rangle \propto z^p$, with  $p= (h-\mu-\nu)/2$.
We represent the solution 
$y$ as a linear combination of $z^p$ from $V_{C}$ and find the
co-efficients. Due to the quadratic dependence of ${\cal H}$ on $E_+$ and $E_-$,
we notice that, the solution space splits as $V_C = V_e \oplus V_o$ --- the even and odd
states of the representation space. The process of finding solutions in $V_e$ and
$V_o$ is demonstrated below, through a set of examples.

{\bf Example 1:} We choose $\gamma=1/2, \ \delta=\varepsilon=-1/2, \
\alpha\beta=1/2$. This leads to $|\alpha-\beta|=1/2$, ensuring the singularities
at $0$ and $\infty$ to be elementary. The Heun equation takes the form
\begin{equation}
\dfrac{d^2y}{dz^2}+
\dfrac{1}{2}\left(\dfrac{1}{z}-\dfrac{1}{z-1}-\dfrac{1}{z-a}\right)
\dfrac{dy}{dz}+\dfrac{z-2q}{2z(z-1)(z-a)}y=0,
\end{equation}
with $\mu=-1, \nu=0$ i.e., $C=-2$. This allows the positive
discrete and negative discrete representation spaces along with a triplet 
representation with $h\in\{-1,0,+1\}$ or $p\in\{0,1/2,1\}$. Choosing the triplet
space, $V_e$ corresponds to $p\in \{0,1\}$ and $V_o$ to
$p\in\{1/2\}$, leading to the solutions
\begin{eqnarray}
y_1 \in V_e&=& z+\sqrt{a}, \quad \mbox{with eigenvalue} \quad q=+\sqrt{a}/2,\\
y_2 \in V_e&=& z-\sqrt{a}, \quad \mbox{with eigenvalue} \quad q=-\sqrt{a}/2 \\
\mbox{and}\quad y_3\in V_o&=& \sqrt{z}, \quad \mbox{with eigenvalue} \quad q=(a+1)/4.\label{ex_one_rootz}
\end{eqnarray}
The positive discrete and negative discrete representations also yield
solutions, not explicitly shown for this example.

{\bf Example 2:} We choose $\gamma=3/2, \ \delta=\varepsilon=-1/2, \
\alpha\beta=0$, leading to $|\alpha-\beta|=1/2$. The Heun equation is given by,
\begin{equation}
\dfrac{d^2y}{dz^2}+
\dfrac{1}{2}\left(\dfrac{3}{z}-\dfrac{1}{z-1}-\dfrac{1}{z-a}\right)
\dfrac{dy}{dz}+\dfrac{q}{z(z-1)(z-a)}y=0
\end{equation}
and corresponds to $\mu=-1/2, \nu=1/2$ with $C=-2$, as in the previous case. Here, for the
triplet space, one obtains $p\in\{-1/2,0,1/2\}$, with $p\in\{0\}$ being $V_o$. The
solutions are:
\begin{eqnarray}
y_1 \in V_e&=& \sqrt{z}+\sqrt{a/z}, \quad \mbox{with eigenvalue} \quad q=-(a+1)/4 +
\sqrt{a}/2,\label{ex_two_rootz1}\\
y_2\in V_e&=& \sqrt{z}-\sqrt{a/z}, \quad \mbox{with eigenvalue} \quad
q=-(a+1)/4 - \sqrt{a}/2\label{ex_two_rootz2}\\
\mbox{and} \quad y_3 \in V_o&=& \mbox{const}., \quad \mbox{with eigenvalue} \quad q=0. 
\end{eqnarray}
Here too, we have positive discrete and negative discrete representations, like
the previous example.

{\bf Example 3:} Next, we analyze the Lam\'{e} equation, which is a special case of the
Heun equation:
\begin{equation}
\dfrac{d^2y}{dz^2}+\dfrac{1}{2}\left(\dfrac{1}{z}+\dfrac{1}{z-1}+
\dfrac{1}{z-a}\right)\dfrac{dy}{dz}-\dfrac{q+\rho(\rho+1)z/4}{z(z-1)(z-a)}y(z)=0.
 \end{equation}
Here, the singularities at $z=0, \ 1$ and $a$ are elementary and choosing 
$\rho(\rho+1)=0$, makes the singularity at $z=\infty$ also elementary. With this
choice, we have $\mu=0=\nu$, i.e., $C=0$, giving us the possibility to choose from
singlet, positive discrete and negative discrete representations. The singlet
solution is, $y_0=\mbox{const}.$, with $q=0$. The positive discrete space splits into the
even $(V_e^+)$ and the odd $(V_o^+)$ subspaces and the corresponding solutions, for a 
chosen eigenvalue $q \in \mathbb{R} $, are:
\begin{eqnarray}
y_1\in V_e^+&=& 1+2 q\left[\dfrac{z}{a}+\dfrac{q+a+1}{3}
\left(\dfrac{z}{a}\right)^2+
\dfrac{2q^2+10 q\left(a+1\right)+8\left(a^2+1\right)+7 a}{45}
\left(\dfrac{z}{a}\right)^3+...\right]\\
y_2\in V_o^+&=&\sqrt{z}\left[1+\dfrac{4q+a+1}{6}
\left(\dfrac{z}{a}\right)+
\dfrac{16q^2+40q\left(a+1\right)+9\left(a^2+1\right)+6a}{120}
\left(\dfrac{z}{a}\right)^2+...\right].\label{positive_odd}
\end{eqnarray}
Similarly, for the negative discrete space too, we have the even $(V_e^-)$ and the odd
$(V_o^-)$ subspaces and the corresponding solutions, for a 
chosen eigenvalue $q \in \mathbb{R} $, are:
\begin{eqnarray}
y_3\in V_e^-&=& 1+2q\left[\dfrac{1}{z}+\dfrac{q+a+1}{3}
\left(\dfrac{1}{z^2}\right)+
\dfrac{2q^2+10q\left(a+1\right)+8\left(a^2+1\right)+7a}{45}
\left(\dfrac{1}{z^3}\right)+...\right]
\\ 
y_4\in V_o^-&=&\dfrac{1}{\sqrt{z}}\left[1+\dfrac{4q+a+1}{6}
\left(\dfrac{1}{z}\right)+
\dfrac{16q^2+40q\left(a+1\right)+9\left(a^2+1\right)+6a}{120}
\left(\dfrac{1}{z^2}\right)+...\right].\label{negative_odd}
\end{eqnarray}
The domain of convergence for the positive discrete space solutions is $0<z<\mbox{min}\{1,a\}$, whereas, that for the negative discrete space solutions is $\mbox{max}\{1,a\}<z<\infty$. The solutions in infinite dimensional spaces can always be written as a power series in $z$ or $1/z$, with an overall factor of $z^p$, $p \in \mathbb{R} $, like the above four solutions. 

For all the above examples, we have obtained solutions, which are \emph{not} available from the previously known $su(1,1)$ symmetry
of the Heun equation~\cite{turbiner}. These solutions are presented here by
 Eqs.~(\ref{ex_one_rootz}), (\ref{ex_two_rootz1}), (\ref{ex_two_rootz2}), (\ref{positive_odd}) and (\ref{negative_odd}). It may be noted that all of them involve powers of $\sqrt{z}$, which reflects the advantage of using a different representation of the $su(1,1)$ generators in the present work.

In conclusion, we have found that, the Heun operator can be exactly cast as a quadratic
polynomial of elements of $su(1,1)$ algebra, if the singularities at $z=0$ and
$z=\infty$ are elementary. This allows one to use the representations of $su(1,1)$ to find explicit solutions of Heun 
equation. The finite dimensional
representations yield polynomial solutions in $\sqrt{z}$, while positive
and negative discrete representations give power series solutions in $z$
and $1/z$, respectively, modulo an overall factor of $z^p$.
Heun equation is connected with a host of physical
problems, related to quantum mechanical and non-linear systems. Hence, the
relevance of the algebraically generated solution space to these problems needs
careful considerations. We intend to return to some of these problems in near
future.

\vspace{2\baselineskip}\vspace{-\parskip}

\textbf{Acknowledgements:} We thank Arunesh Roy for useful discussions.

\vspace{1\baselineskip}\vspace{-\parskip}

\end{document}